\documentstyle[prl,twocolumn,aps]{revtex}
\RequirePackage{epsfig}
\def\Journal#1#2#3#4{{#1} {\bf #2}, #3 (#4)}
\def\PR{\em Phys. Rev.}
\def\PRL{\em Phys. Rev. Lett.}
\def\PRA{{\em Phys. Rev.} A}
\def\JMP{\em J. Math. Phys.}

\def\JETP{\em Soviet Phys. JETP}
\def\Science{\em Science}
\def\JPB{\em J. Phys. B}
\begin{document}
\draft
\title {Many-body solitons in a one-dimensional condensate
        of hard core bosons}
\author{M. D. Girardeau$^{1,2}$ and E. M. Wright$^2$}
\address{$^1$Institute of Theoretical Science, University of Oregon, 
Eugene, OR 97403\\
	$^2$Optical Sciences Center and Department of Physics,
University of Arizona, Tucson, AZ 85721}
 \date{\today}
\maketitle
\begin{abstract}
A mapping theorem leading to exact many-body dynamics of impenetrable bosons
in one dimension reveals dark and gray soliton-like structures in a
toroidal trap which is phase-imprinted.  On long time scales revivals appear
that are beyond the usual mean-field theory.
\end{abstract}
\pacs{03.75.Fi,03.75.-b,05.30.Jp}
Dark and gray solitons are a generic feature
of the nonlinear Schr\"{o}dinger
equation with repulsive interactions, and several calculations of
their dynamics
based on the Gross-Pitaevskii (GP) equation have appeared
\cite{Reinhardt,Dum,Scott,Jackson,Burger,Denschlag,Muryshev,Busch},
as well as experiments demonstrating their existence in atomic BECs
\cite{Burger,Denschlag}.
Since the GP equation is a nonlinear approximation to the more exact 
{\em linear} many-body Schr\"{o}dinger equation, this raises the question of 
how observed solitonic behavior arises in a theory which is linear at a
fundamental level. Here this issue will be examined with the aid of
exact many-body solutions. It has been shown by Olshanii \cite{Olshanii} that 
at sufficiently low temperatures, densities, and
large positive scattering length, a BEC in a thin atom waveguide
has dynamics which approach those of
a one-dimensional (1D) gas of impenetrable point bosons. This is
a model for which the exact many-body energy eigensolutions were
found in 1960 using an exact mapping from the Hilbert space of energy
eigenstates of an {\em ideal} gas of fictitious spinless fermions to that of 
many-body eigenstates of impenetrable, and therefore 
{\em strongly interacting}, bosons \cite{map,map2}. 
The term ``Bose-Einstein condensation" is used here in a
generalized sense; it was shown by Lenard \cite{Lenard} and by Yang and 
Yang \cite{Yang} that for the many-boson ground state of this system, the
occupation of the lowest single-particle state is of order $\sqrt{N}$ where 
$N$ is the total number of atoms, in contrast to $N$ for usual BEC.
Nevertheless, since $N\gg 1$ and the momentum distribution has a sharp peak
in the neighborhood of zero momentum \cite{Olshanii}, this system shows strong
coherence effects typical of BEC. The response of a BEC of this type to 
application of a delta-pulsed optical lattice was recently calculated by 
Rojo {\it et al.} \cite{Rojo}, using the Fermi-Bose mapping theorem 
\cite{map,map2}, as an exactly calculable model of dynamical optical lattice 
behavior. They found spatial focussing and periodic self-imaging (Talbot 
effect), which decay as a result of interactions. This decay is absent in the 
GP approximation and therefore serves as a signature of many-body interaction 
effects omitted in GP.

In this Letter we examine the appearence of dark soliton-like structures
using the model of a 1D hard-core Bose gas
in a toroidal trap, or ring, with cross section so small
that motion is essentially circumferencial
\cite{JavPaiYoo98,SalParRea99,BusAng99,BenRagSme97,Rok}.  The Fermi-Bose
mapping is employed to generate exact solutions for this problem.
We identify
stationary solutions which reflect some properties of dark solitons
from the GP theory when the ring is pierced at a point by an intense
blue-detuned laser.  We also present dynamical solutions when half of
an initially homogeneous ring BEC is phase-imprinted via the light-shift
potential of an applied laser, leading to gray soliton-like
structures whose velocity depends on the imposed phase-shift
\cite{Burger,Denschlag}.  Such structures are apparent for times less than
the echo time $\tau_e=L/c$, with $L$ the ring circumference and $c$ the
speed of sound in the BEC.  On longer time scales the dynamics becomes
very complex showing Talbot recurrences which are beyond
the GP theory.

{\it Time-dependent Fermi-Bose mapping theorem:}
The original proof
\cite{map,map2} was restricted to energy eigenstates, but the generalization
to the time-dependent case is almost trivial. 
The Schr\"{o}dinger Hamiltonian is assumed to have the structure
\begin{equation}\label{eq1}
\hat{H}=\sum_{j=1}^{N}-\frac{\hbar^2}{2m}\frac{\partial^2}{\partial x_{j}^{2}}
+V(x_{1},\cdots,x_{N};t)  ,
\end{equation}
where $x_j$ is the one-dimensional position of the $j{\it th}$ particle
and $V$ is symmetric (invariant) under permutations of the particles.
The two-particle interaction 
potential is assumed to contain a hard core of 1D diameter $a$. This is
conveniently treated as a constraint on allowed wave functions
$\psi(x_{1},\cdots,x_{N};t)$:
\begin{equation}\label{eq2}
\psi=0\quad\text{if}\quad |x_{j}-x_{k}|<a\quad,\quad 1\le j<k\le N  ,
\end{equation}
rather than as an infinite contribution to $V$, which then consists of all
other (finite) interactions and external potentials. The time-dependent 
version starts from fermionic solutions $\psi_{F}(x_{1},\cdots,x_{N};t)$ of 
the time-dependent many-body Schr\"{o}dinger equation (TDMBSE) 
$\hat{H}\psi=i\hbar\partial\psi/\partial t$ which are antisymmetric under
all particle pair exchanges $x_{j}\leftrightarrow x_{k}$, hence all
permutations. As in the original theorem \cite{map},
define a ``unit antisymmetric function"
\begin{equation}\label{eq3}
A(x_{1},\cdots,x_{N})=\prod_{1\le j<k\le N}\text{sgn}(x_{k}-x_{j})  ,
\end{equation}
where $\text{sgn}(x)$ is the algebraic sign of the coordinate difference
$x=x_{k}-x_{j}$, i.e., it is +1(-1) if $x>0$($x<0$). For given 
antisymmetric $\psi_F$,
define a bosonic wave function $\psi_B$ by
\begin{equation}\label{eq4}
\psi_{B}(x_{1},\cdots,x_{N};t)=A(x_{1},\cdots,x_{N})\psi_{F}(x_{1},\cdots,
x_{N};t)
\end{equation}
which defines the Fermi-Bose mapping. $\psi_B$ satisfies
the hard core constraint (2) if $\psi_F$ does, is totally
symmetric (bosonic) under permutations, obeys the same
boundary conditions as $\psi_F$, {\it e.g.} periodic boundary conditions
on a ring, and
$\hat{H}\psi_{B}=i\hbar\partial\psi_{B}/\partial t$ follows from
$\hat{H}\psi_{F}=i\hbar\partial\psi_{F}/\partial t$ \cite{map,map2}.

{\it Exact solutions for impenetrable point bosons:} The mapping theorem
leads to explicit expressions for all many-body energy eigenstates and
eigenvalues of a 1D scalar condensate (bosons all of the same spin) 
under the assumption that the only two-particle interaction is a
zero-range hard core repulsion, represented by the $a\rightarrow 0$
limit of the hard-core constraint. Such solutions were obtained in 
Sec. 3 of the original work \cite{map} for periodic
boundary conditions and no external potential. The exact many body
ground state was found to be a pair product of Bijl-Jastrow form: 
$\psi_{0}=\text{const.}\prod_{i>j}|\sin[\pi L^{-1}(x_{i}-x_{j})]|$.
In spite of the very long range
of the individual pair correlation factors $|\sin[\pi L^{-1}(x_{i}-x_{j})]|$,
the pair distribution function $D(x_{ij})$, the integral of $|\psi_{0}|^2$
over all but two coordinates, was found to be of short range:
$D(x_{ij})=1-j_0^2(\pi\rho x_{ij})$, with
$j_{0}(\xi)=\sin\xi/\xi$, the spherical Bessel function of order zero.
The system was
found to support propagation of sound with speed $c=\pi\hbar\rho/m$ where
$\rho=N/L$, the 1D atom number density. 

To generalize to the time-dependent case, assuming that 
the many-body potential of Eq. (1) is a sum of one-body 
external potentials $V(x_{j},t)$, one generalizes the time-independent 
determinantal many-fermion wavefunction \cite{map} to a determinant 
\begin{equation}\label{eq5}
\psi_{F}(x_{1},\cdots,x_{N};t)=C\det_{i,j=1}^{N}\phi_{i}(x_{j},t)  ,
\end{equation}
of solutions $\phi_{i}(x,t)$ of the {\em one-body} TDSE in the external 
potential $V(x,t)$. It then follows that $\psi_F$ satisfies the TDMBSE, 
and it satisfies the impenetrability constraint (vanishing when any 
$x_{j}=x_{\ell}$) trivially due to antisymmetry. Then by the mapping theorem 
$\psi_B$ of Eq.(4) satisfies the same TDMBSE.

{\it Dark solitons on a ring:} Consider $N$ bosons in a tight toroidal trap,
and denote their 1D positions measured around the circumference by
$x_j$. This is equivalent to the exactly-solved model \cite{map} of
$N$ impenetrable point bosons in 1D with wave functions satisfying periodic
boundary conditions with period $L$ equal to the torus circumference, and
the fundamental periodicity cell may be chosen as $-L/2<x_{j}<L/2$.
However, the rotationally invariant quantum states of this problem do not
reveal any dark soliton-like structures.  To proceed we therefore
consider the case that a blue-detuned laser field pierces the ring
at $x=0$ by virtue of the associated repulsive dipole force: The light
sheet then provides a reference position for the null of the dark
soliton.  Assume that the
light sheet is so intense and narrow that it may be replaced by a
constraint that the many-body wave function (hence the orbitals
$\phi_i$) must vanish whenever any $x_{j}=0$. Then
the appropriate orbitals $\phi_{i}(x)$ are free-particle
energy eigenstates vanishing at $x=0$ and periodic with period $L$.
The complete orthonormal set of even-parity eigenstates $\phi_{n}^{(+)}$
and odd-parity eigenstates $\phi_{n}^{(-)}$ are
%Eq. (6)
\begin{eqnarray}
\phi_{n}^{(+)}(x) & = & \sqrt{2/L}\sin[(2n-1)\pi|x|/L]  , \nonumber\\
\phi_{n}^{(-)}(x) & = & \sqrt{2/L}\sin(2n\pi x/L)       ,
\end{eqnarray}
with $n$ running from $1$ to $\infty$. The odd eigenstates are
the same as those of free particles with no $x=0$ constraint, since these
already vanish at $x=0$. However, the even ones are strongly affected by the
constraint, their cusp at $x=0$ being a result of the impenetrable light sheet
at that point. If one bends a 1D box $-L/2<x<L/2$ with impenetrable 
walls into a ring, identifying the walls at $\pm L/2$, then those
particle-in-a-box eigenfunctions which are even about the box center become
identical with the $\phi_{n}^{(+)}$, and their cusp results from
the nonzero slope of these functions at the walls. The $N$-fermion ground state
is obtained by inserting the lowest $N$ orbitals (6) into the determinant (5) 
(filled Fermi sea). Assume that $N$ is odd.
Since $\phi_{1}^{(+)}$ is lower than $\phi_{1}^{(-)}$, this Fermi sea 
consists of the first $(N+1)/2$ of the $\phi_{n}^{(+)}$
and the first $(N-1)/2$ of the $\phi_{n}^{(-)}$. The $N$-boson
ground state is then given by (4). Since $A^{2}=1$, its one-particle 
density $\rho(x)$ is the same as that of the $N$-fermion ground state,
the sum of partial densities contributed by all one-particle states in the 
Fermi sea. Thus it is the sum of
%Eq. (7)
\begin {equation}
\rho^{(+)}(x)
=\frac{N+1}{2L}-\frac{\sin[2(N+1)\pi x/L]}{2L\sin(2\pi x/L)}  ,
\end{equation}
and
%Eq. (8)
\begin{equation}
\rho^{(-)}(x)
=\frac{N-1}{2L}-\frac{\sin[(N-1)\pi x/L]\cos[(N-3)\pi x/L]}
{L\sin(2\pi x/L)}
\end{equation}
In the thermodynamic limit
$N\rightarrow\infty$, $L\rightarrow\infty$, $N/L\rightarrow\rho$ for fixed 
$x$, $\rho^{(\pm)}$ each contribute half of the total density
$\rho(x)$:
%9
\begin{equation}
\rho(x)\sim \rho[1-j_{0}(2\pi\rho x)]  .
\end{equation}
Since $j_{0}(0)=1$, $\rho(x)$ vanishes at $x=0$ and approaches the mean
density $\rho$ over a healing length $L_h=1/2\rho$ with damped spatial
oscillations about its limiting value.  This differs in detail from the
density $\rho_\infty\tanh^{2}(x/w)$ of a GP dark soliton \cite{Gin},
with $\rho_\infty$ the background density and $w$ the corresponding
healing length,
but has some qualitative similarity.  However, it is only the odd
component $\rho^{(-)}(x)\approx\rho(x)/2$ which has the feature of a dark
soliton that the corresponding odd orbitals have a $\pi$ phase-jump at $x=0$
(and also at $x=\pm L/2$ to obey the periodic boundary conditions).  But the
odd and even components can never be separated physically, so the
odd dark soliton-like component is always accompanied by the even
non-soliton component.

Next, suppose that the light-sheet is turned off at $t=0$ by removing
the constraint that the wave function vanish at $x=0$.
The solution of the TDMBSB for the many-boson system is then given by
(4) where the Slater determinant (5) is built from the first $(N+1)/2$
of the $\phi_{n}^{(+)}(x,t)$ and the first $(N-1)/2$ of the 
$\phi_{n}^{(-)}(x,t)$, where these time-dependent orbitals are solutions
of the single-free-particle TDSE which (a) reduce to the orbitals (6) at 
$t=0$, and (b) satisfy periodic boundary conditions with periodicity cell
$-L/2<x<L/2$. The odd solutions are trivial: Since these never
``see" the $x=0$ constraint even for $t<0$, they differ from the odd orbitals
(6) only by time-dependent phase shifts: 
$\phi_{n}^{(-)}(x,t)=\phi_{n}^{(-)}(x)e^{-i\omega_{n}t}$ with 
$\omega_{n}=\hbar k_{n}^{2}/2m$ and $k_{n}=2n\pi/L$. It
follows that $\rho^{(-)}(x,t)$ is time-independent, and given in the
thermodynamic limit by
%10
\begin{equation}
\rho^{(-)}(x,t)\sim (\rho/2)[1-j_{0}(2\pi\rho x)]  .
\end{equation}
This further reinforces our view that the odd component of the density
shares features of a dark soliton. The even-parity orbitals
$\phi_{n}^{(+)}(x,t)$ are complicated since the
removal of the light sheet constitutes a large, sudden perturbation. Indeed,
the periodic even-parity solutions of the free-particle 
Schr\"odinger equation are 
$\chi_{p}^{(+)}(x)=\sqrt{(2-\delta_{p0})/L}\cos(2p\pi x/L)$
with $p=0,1,2,\cdots$, and these are very different from the solutions
$\phi_{n}^{(+)}(x)$ with the $x=0$ constraint [Eq. (6)]. Nevertheless, since
the $\chi_{p}^{(+)}(x)$ are complete for the subspace of even-parity, spatially
periodic functions, one can expand the $\phi_{n}^{(+)}(x,t)$ in terms of the
$\chi_{p}^{(+)}(x)$, which evolve with time-dependent phases 
$e^{-i\omega_{p}t}$ with $\omega_{p}=\hbar k_{p}^{2}/2m$ and
$k_{p}=2p\pi/L$. One finds 
%11
\begin{equation}
\phi_{n}^{(+)}(x,t)=\frac{2(2n-1)}{\pi}\sqrt{\frac{2}{L}}
\sum_{p=0}^{\infty}\frac{(2-\delta_{p0})\cos(k_{p}x)e^{-i\omega_{p}t}}
{(2n-1)^{2}-4p^{2}}
\end{equation}
$\rho^{(+)}(x,t)$ is the sum of absolute squares of the first
$(N+1)/2$ of the sums (11), generalizing (7). Adding the time-independent 
expression $\rho^{(-)}(x,t)$, given in the thermodynamic limit by (10) or 
exactly by (8), one finds the time-dependent total density $\rho(x,t)$.
There are two important time scales: One is the Poincar\'{e}
recurrence time $\tau_r$. Noting that $\omega_p$ in (11) is proportional to
$p^2$, one finds that all terms in the sum are time-periodic with period
$\tau_{r}=mL^{2}/\pi\hbar$, which is therefore the recurrence time for the
density and in fact all properties of our model \cite{Rojo}.
The other important time is the echo
time $\tau_e$, the time for sound to make one circuit around the torus.
Recalling
that the speed of sound in this system is $c=\pi\hbar\rho/m$ \cite{map},
one finds $\tau_{e}=\tau_{r}/N$.  For $t<<\tau_e$ after the constraint is
removed, the initial density develops sound waves that propagate around
the ring, and that we examine below in the context of phase-imprinting.
For $t>\tau_e$ the evolution is very complex, but complete recurrences
occur for times $t=n\tau_r$ with fractional revivals in between.

\begin{figure}
\epsfxsize 3in
\epsfbox{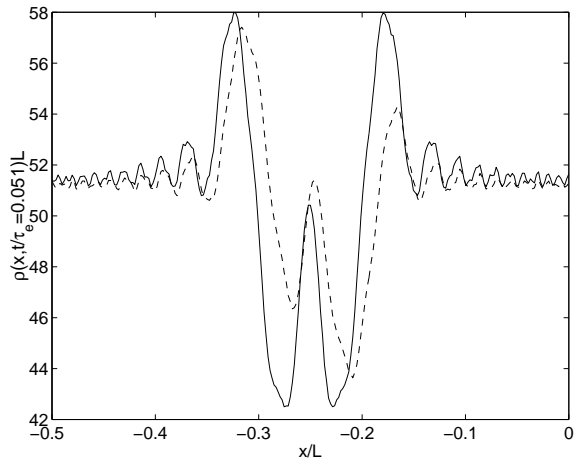}
\caption{Scaled density $\rho(x,t)L$ versus scaled position around the
ring $x/L$ for $N=51$, $t/\tau_e=0.051$, and $\Delta\theta=\pi$
(solid line), and $\Delta\theta=0.5\pi$ (dashed line).
Due to symmetry we show only half
of the ring $-L/2<x<0$, the phase-jump being imposed at $x=-L/4$.}
\label{Fig:one}
\end{figure}

{\it Gray soliton formation by phase-imprinting:} Consider next
a toroidal BEC in its ground state to which a phase-imprinting
laser is applied over half the ring at $t=0$.  This is
described by the single-particle Hamiltonian
%12
\begin{equation}
\hat{H}=\sum_{j=1}^{N}\left[-\frac{\hbar^2}{2m}\frac{\partial^2}
{\partial x_{j}^{2}}-\hbar\Delta\theta\delta(t)S(x_{j})\right]
\end{equation}
where $S(x)=\theta(L/4-|x|)$, i.e., unity for $-L/4<x<L/4$ and
zero elsewhere. This is the technique used in recent experiments
\cite{Burger,Denschlag}, here idealized to
a delta-function in time and to sharp spatial edges. Before the pulse the
most convenient free-particle orbitals in (5)
are plane waves $\phi_{n}(x)=\sqrt{(1/L)}e^{ik_{n}x}$ where $k_{n}=2n\pi/L$
and $n=-n_{F},-n_{F}+1,\cdots,n_{F}-1,n_{F}$ with $n_{F}=(N-1)/2$. Let
$\phi_{n}(x,t)$ be the solution
of the TDSE with the Hamiltonian (12) reducing to the above
$\phi_{n}(x)$ just before the pulse. Then the solutions just
after the pulse are $\phi_{n}(x,0+)=\phi_{n}(x)e^{iS(x)\Delta\theta}$.  The
potential gradients at the pulse edges impart momentum kicks to the particles
there which induce both compressional waves propagating at the speed, $c$, 
of sound and density dips (gray solitons) moving at speeds $|v|<c$.
The expansion of $\phi_{n}(x,t)$ in terms of the unperturbed plane waves is
evaluated as
%Eq.(13)
\begin{eqnarray}
\phi_{n}(x,t) & = & \frac{1}{2}\left(1+e^{i\Delta\theta}\right)
-\frac{1-e^{i\Delta\theta}}{\pi}
\sum_{\ell=-\infty}^{\infty}\nonumber\\
& \times & \frac{(-1)^{\ell}\phi_{n-2\ell-1}(x)
e^{-i\omega_{n-2\ell-1}t}}{2\ell+1}
\end{eqnarray}
and the time-dependent density is the sum of the absolute squares of the
lowest $N$ of these. Figure \ref{Fig:one} shows numerical simulations obtained
using Eq. (13) for $N=51$, $t/\tau_e=0.051$, and
$\Delta\theta=\pi$ (solid line), and $\Delta\theta=0.5\pi$
(dashed line): due to symmetry we show only half
of the ring $-L/2<x<0$, the phase-shift being imposed at $x=-L/4$.
Considering times short compared to the echo time means that the
corresponding results are not very sensitive to the periodic
boundary conditions, and also therefore apply to a linear geometry.
The initial density profile is flat with a value $\rho_0 L=51$.
For both phase-shifts two distinct maxima are seen,
which travel at close to the
speed of sound $c$, and two distinct minima, which are analogous to
gray solitons and travel at velocities $|v|/c<1$. 
In addition, there are
also high wavevector oscillations which radiate at velocities greater than
$c$.  In the case of a phase-shift
$\Delta\theta=\pi$, the density is symmetric about $x=-L/4$,
whereas for a phase-shift other than a multiple of $\pi$ the evolution is
not symmetric, see the dashed line where the global minimum moves to the
right in reponse to the phase-shift. 
\begin{figure}
\epsfxsize 3in
\epsfbox{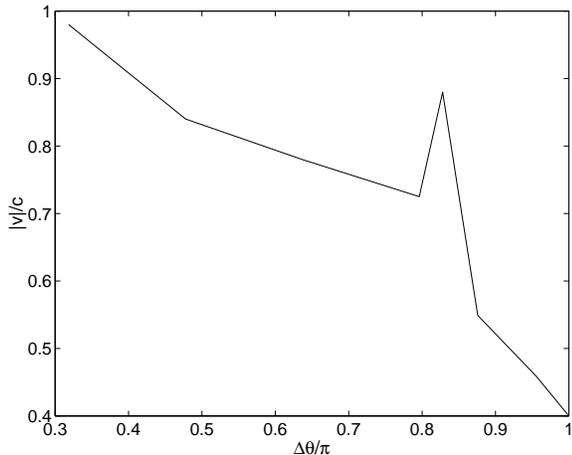}
\caption{Dark soliton velocity $|v|/c$ scaled to the speed of sound
$c$ as a function of phase-shift $\Delta\theta/\pi$ for $N=51$.}
\label{Fig:two}
\end{figure}
In Fig. \ref{Fig:two} we plot the 
calculated
velocity of the global density minimum relative to the speed of sound
for a variety of phase-shifts $\Delta\theta$.  The basic trend is that
larger phase-shift means lower velocity, in qualitative agreement with
recent experiments \cite{Burger,Denschlag}, but there is a sharp velocity
peak at $\Delta\theta\approx 0.83\pi$: This peak results from the cross-over
between two local minima in the density.  These general features,
the generation of gray solitons
and density waves, agree with those of the GP theory, but here arise out of
the exact many-body calculation.

In conclusion, using our exactly-soluble 1D model we hope to have shown
that the dark solitonic features of atomic BECs normally described within
the mean-field GP theory arise naturally from consideration of the
exact {\it linear} many-body theory for times less than the echo time.
An advantage of this approach is
that it is number-conserving and does not rely on any symmetry-breaking
approximation.  In addition, long time dynamics such as collapses and
revivals are accounted for \cite{Rojo}.  A detailed comparison between
our results and current experiments is not possible
as they do not conform to the conditions
for a 1D system.  However, some estimates are in order to set the
appropriate time scales: If we consider $^{87}$Rb with a ring of
circumference $L=100$ $\mu$m, and a high transverse trapping frequency
$\omega_\perp=2\pi\times 10^5$ Hz, then we are limited to atom
numbers $N<300$ \cite{Olshanii}, so these are small condensates.
We then obtain $\tau_r=4.6$ s, and $\tau_e=90$ ms for $N=51$.
Finally, we remark that since our approach relied on the mapping
between the strongly-interacting Bose system and a non-interacting
``spinless Fermi gas" model, this suggests that dark and gray solitons
should also manifest themselves in the density for the 1D Fermi system.
Although real fermions have spin,
the interactions used here to generate solitons were spin-independent.
\vspace{0.2cm}

\noindent
This work was supported by the Office of Naval Research Contract
No. N00014-99-1-0806.

\end{document}